\documentclass[12pt, preprint]{aastex}
\usepackage{lscape}
\slugcomment{To appear in ApJ}

\shorttitle{Gemini Imaging of Mid-IR Emission from the Nuclear Region of Centaurus A}
\shortauthors{Radomski et al.}

\begin{document}

\title{Gemini Imaging of Mid-IR Emission from the Nuclear Region
of Centaurus A}

%% Use \author, \affil, and the \and command to format
%% author and affiliation information.
%% Note that \email has replaced the old \authoremail command
%% from AASTeX v4.0. You can use \email to mark an email address
%% anywhere in the paper, not just in the front matter.
%% As in the title, use \\ to force line breaks.

\author{James T. Radomski\altaffilmark{1}, Christopher Packham\altaffilmark{2},
N. A. Levenson\altaffilmark{3}, Eric Perlman\altaffilmark{4}, Lerothodi L. Leeuw\altaffilmark{5}, Henry
Matthews\altaffilmark{6}, Rachel Mason\altaffilmark{7}, James M. De Buizer\altaffilmark{1}, Charles M.
Telesco\altaffilmark{2}, Manuel Orduna\altaffilmark{2}}

%% Mark off your abstract in the ``abstract'' environment. In the manuscript
%% style, abstract will output a Received/Accepted line after the
%% title and affiliation information. No date will appear since the author
%% does not have this information. The dates will be filled in by the
%% editorial office after submission.

\begin{abstract}
We present high spatial resolution mid-IR images of the nuclear region of NGC 5128 (Centaurus A).
Images were obtained at 8.8$\micron$, N-band (10.4$\micron$), and 18.3$\micron$ using the mid-IR
imager/spectrometer T-ReCS on Gemini South. These images show a bright unresolved core surrounded by
low-level extended emission. We place an upper limit to the size of the unresolved nucleus of 3.2 pc
(0$\farcs$19) at 8.8$\micron$ and 3.5 pc (0$\farcs$21) at 18.3$\micron$ at the level of the FWHM. The
most likely source of nuclear mid-IR emission is from a dusty torus and possibly dusty narrow line
region with some contribution from synchrotron emission associated with the jet as well as relatively
minor starburst activity. Clumpy tori models are presented which predict the mid-IR size of this torus
to be no larger than 0$\farcs$05 (0.85pc). Surrounding the nucleus is extensive low-level mid-IR
emission. Previously observed by ISO and Spitzer, this paper presents to date the highest spatial
resolution mid-IR images of this extended near nuclear structure. Much of the emission is coincident
with Pa-$\alpha$ sources seen by HST implying emission from star forming areas, however evidence for
jet induced star formation, synchrotron emission from the jet, a nuclear bar/ring, and an extended
dusty narrow emission line region is also discussed.
\end{abstract}

%% Keywords should appear after the \end{abstract} command. The uncommented
%% example has been keyed in ApJ style. See the instructions to authors
%% for the journal to which you are submitting your paper to determine
%% what keyword punctuation is appropriate.

%% Authors who wish to have the most important objects in their paper
%% linked in the electronic edition to a data center may do so in the
%% subject header.  Objects should be in the appropriate "individual"
%% headers (e.g. quasars: individual, stars: individual, etc.) with the
%% additional provision that the total number of headers, including each
%% individual object, not exceed six.  The \objectname{} macro, and its
%% alias \object{}, is used to mark each object.  The macro takes the object
%% name as its primary argument.  This name will appear in the paper
%% and serve as the link's anchor in the electronic edition if the name
%% is recognized by the data centers.  The macro also takes an optional
%% argument in parentheses in cases where the data center identification
%% differs from what is to be printed in the paper.

\keywords{galaxies: active - galaxies: nuclei - infrared: galaxies - galaxies: individual (Cen A) -
galaxies: Seyfert}

\altaffiltext{1}{Gemini South Observatory, Casilla 603, La Serena, Chile}

\altaffiltext{2}{Astronomy Department, University of Florida, 211 Bryant Space Science Center, P.O. Box
112055, Gainesville, FL 32611-2055}

\altaffiltext{3}{Department of Physics and Astronomy, University of Kentucky, Lexington, KY 40506}

\altaffiltext{4}{Joint Center for Astrophysics, Department of Physics, University of Maryland,
Baltimore County, 1000 Hilltop Circle, Baltimore, MD 21250}

\altaffiltext{5}{Department of Physics and Electronics, Rhodes University, PO Box 94, Grahamstown 6140,
South Africa.}

\altaffiltext{6}{Herzberg Institute of Astrophysics, National Research Council, P.O. Box 248,
Penticton, BC, V2A 6J9}

\altaffiltext{7}{Gemini North Observatory, 670 N. A'ohoku Place, Hilo, HI  96720}

\section{Introduction}

Centaurus A (NGC 5128, hereafter Cen A) is a nearby elliptical galaxy and the prototypical
Faranoff-Riley Class I source with spectacular twin radio lobes extending out to $\sim$ 4$\degr$
($\sim$ 250 kpc) from the nucleus (see Israel 1998 for an excellent review). The morphology as well as
a bimodal metallicity distribution of the globular cluster population indicate the galaxy has
experienced a major merger event in the past (perhaps as recently as 1.6 $\times$ 10$^{8}$ yr ago;
Quillen et al. 1993). On nuclear scales, evidence for a central super massive black hole (SMBH) comes
from variable x-ray and radio observations. Estimates of the black hole mass by Silge et al. (2005),
Marconi et al. (2006), and Haring-Neumayer et al. (2006) suggest it to be $\sim$10$^{8}$ M$_{\sun}$.

Despite intensive studies of Cen A, observations at UV and optical wavelengths have been especially
hampered by the dust lane that bisects the galaxy and heavily obscures the nuclear regions.
Additionally, unified theories of AGN (Krolik \& Begelman 1988; Antonucci 1993: Urry \& Padovani 1995),
suggest the central SMBH is obscured from our line-of-sight by a geometrically and optically thick
torus of gas and dust that further extinguishes radiation arising from accretion onto the SMBH. High
spatial resolution mid-IR observations now available with 8m class telescopes such as Gemini offer an
ideal way to study the obscured central region of Cen A, having extinction 25-75 times lower than
observations at optical wavelengths. High spatial resolution mid-IR imaging of other nearby AGNs (e.g.
Radomski et al. 2002, 2003, 2003b; Packham et al. 2005, Soifer et al. 2000, 2001, 2003; Perlman et al.
2001, Mason et al. 2007, Alonso-Herrero et al. 2006) highlight the value of such observations in
separating and modeling emission from nuclear star formation regions, the torus, and dusty narrow
emission line regions.

 Previous mid-IR observations of Cen A have produced somewhat contradictory
results. Thus, the true nature and extent of the nuclear emission remains controversial. Observations
by Whysong \& Antonucci (2004) on the Keck I detected an unresolved nucleus at 11.7$\micron$ and
17.75$\micron$ at resolutions of $\sim$ 0$\farcs$3 and 0$\farcs$5 respectively, placing limits of 5.6pc
and 8.85pc on the size of the mid-IR emitting region. However, Karovska et al. (2003) (hereafter K03)
claim to resolve the nucleus of Cen A based on 8.8$\micron$ and N-band imaging with the Magellan 6.5m
telescope, calculating a size of 0$\farcs$17 $\pm$ 0$\farcs$02 or 3pc after subtracting a PSF in
quadrature. Further observations by Siebenmorgen, Kr{\"u}gel, \& Spoon (2004) with a resolution of
$\sim$ 0$\farcs$5 were unable to directly resolve the nucleus at 10.4$\micron$, but showed that the
growth curve indicated weak extended emission over 2$\arcsec$ with a surface brightness a factor of ten
lower than the peak of the nucleus. Most recently, observations by Hardcastle, Kraft, and Worrall
(2006) (herein HKW06) find the nuclear FWHM of Cen A to be slightly resolved calculating a conservative
upper limit of 0$\farcs$27 or $\leq$ 4.6pc at N-band ($\lambda$$_0$ = 10.4$\micron$). If the nucleus of
Cen A is resolved, this could be the first galaxy in which the emission from dust associated with the
torus has been resolved with single-telescope mid-IR observations. In order to address the controversy
of the central nuclear resolution of Cen A, we have methodically observed Cen A and corresponding
calibration sources at high-resolution in the mid-IR to (1) test the hypothesis that the nucleus is
resolved and (2) characterize the mid-IR emission mechanisms. In addition, we draw upon publicly
accessible archived observations in the mid-IR in addition to our own data to further explore low-level
extended emission surrounding the nucleus of Cen A. Our observations are discussed in $\S$ 2 while
results and analysis are presented in $\S$ 3 and $\S$ 4 respectively. Throughout this paper we assume a
distance to Cen A of D= 3.5 Mpc (1$\arcsec$=17pc).

\section{Observations and Data Reduction}

Observations of Cen A were made over two epochs. Epoch A occurred on 2004 January 28 and 29 (UT) and
was initiated by the authors of this paper. Epoch B was observed on 2004 March 6, 11, and 12 and was
retrieved from the Gemini Science Archive (GSA) based on data first published by HKW06. Both sets of
data were obtained using the facility mid-IR camera/spectrograph T-ReCS (Telesco et al. 1998) on the
Gemini South telescope. T-ReCS uses a Raytheon 320 x 240 pixel Si:As IBC array, providing a plate scale
of 0$\farcs$089 pixel$^{-1}$, corresponding to a field of view of 28$\farcs$5 x 21$\farcs$4. The
detector was read out in correlated quadruple sampling (CQS) mode (Sako et al 2003). The standard
chop-nod technique was used to remove time-variable sky background, telescope thermal emission, and the
so-called ``1/f" detector noise. The chop throw was 15$\arcsec$ and the telescope was nodded every 30
sec. All data were reduced using IDL.

\subsection{Epoch A 2004 January 28 and 29}

We carefully observed Cen A over a period of two nights. Images were obtained in the Si-2
($\lambda$$_0$ = 8.74$\mu$m, $\Delta$$\lambda$ = 0.78$\micron$) and Qa ($\lambda$$_0$ = 18.3$\micron$,
$\Delta$$\lambda$ = 1.5$\micron$) filters to optimize resolution and sensitivity in both the 10 and
20$\micron$ atmospheric windows. The chop and nod direction was fixed at 0$\degr$ (North-South). The
first night we observed at 8.8$\micron$ for about 3.5 hours real-time
(real-time$\sim$3.5$\times$on-source time due to chop-nod procedure, not including time for telescope
slewing and calibration) using an iterative procedure which repeatedly interleaved observations of a
point spread function (PSF) star (65 sec on-source) with those of Cen A (455 sec on-source). This
resulted in a total of 5 PSF observations (5.4 min on-source total) and 5 galaxy observations (34 min
on-source total). We used the same procedure on the second night at 18.3$\micron$; however, the initial
PSF star had too low a S/N ratio so a new PSF was chosen resulting in only 3 high S/N PSF measurements
(3.3 min on-source total). A total of 4 galaxy observations (26.4 min on-source total) at 18.3$\micron$
were obtained. Estimates of the FWHM of the galaxy and PSF were done by fitting a Moffat function
(Moffat 1969).

 The stars PPM318494 and PPM291667, located 1.5$\degr$ and 4.2$\degr$ from Cen
 A, were used for PSF comparison to look for extended emission in the galaxy.
 Each of the PSF observations was made immediately prior to or
after the Cen A observations and using an identical observational setup to accurately sample the
delivered image quality of those observations. PSF images were not rotated to correct for changes in
the pupil due to the small rotation angle of the pupil between observations of the galaxy and PSF
(typically $<$ 6$\degr$). This was done to avoid the slight smoothing of the FWHM that occurs when an
image is digitally rotated to correct for pupil rotation. Line cuts taken at similar pupil angles on
the galaxy justify this. These showed the profiles to be radially symmetric hence an azimuthally
averaged FWHM was adequate for investigating any resolved emission in the nucleus of Cen A.

Observations of HD110458 were used for flux calibration through both Si-2 and Qa filters at a similar
airmass as the Cen A observations. Absolute calibration was achieved using the TIMMI2 photometric
standard list where the flux densities of HD 110458 at 8.8 $\micron$ and 18.3 $\micron$ are 6.95 Jy and
1.72 Jy respectively. Observations of HD110458 and Cen A at 8.8 $\micron$ show variations of $\lesssim
5$\%. At 18.3 $\micron$ only one observation of the flux calibrator was taken. However, observations of
the relative flux of Cen A at 18.3 $\micron$ showed variations of $\sim$ 25\%.

The differences in the intrinsic spectra of the stellar sources and Cen A as observed through the Si-2
and Qa filters requires a color correction. This color correction not only affects the observed fluxes
but also the FWHM. A stellar spectra peaking at shorter wavelengths (T$\sim$4000K) will dominate the
shorter wavelength portion of the filter. A cooler source such as Cen A (T$\sim$200K) will peak at
longer wavelengths and dominate the longer wavelength portion of the filter. The resultant effect will
not only alter the flux calibration but also cause the FWHM of ``hot" stellar sources to be smaller
than the relatively ``cool" Cen A when observed through the same filter. This effect is significant in
large bandpass mid-IR filters such as N-band ($\lambda$$_0$ = 10.4$\mu$m, $\Delta$$\lambda$ =
5.3$\micron$). The difference between a T$\sim$4000K and T$\sim$200K source in the N-band will result
in a color correction of the observed flux density up to 21\% and the observed FWHM of the stellar
source will appear 12\% smaller. Due to the relatively small bandwith of the Si-2 and Qa filters
($\Delta$$\lambda$ = 0.78 and 1.5$\micron$) this effect is much smaller. The color correction between a
T$\sim$4000K and T$\sim$200K source is $<$ 4\% and the difference in the FWHM is $<$ 0.05\%, both
within the errors of our observations.

\subsection{Epoch B 2004 March 6, 11, and 12}

Epoch B data were first published by HKW06 and was retrieved from the GSA after its proprietary period
to compare with our observations. Images were obtained in the N-band ($\lambda$$_0$ = 10.4$\mu$m,
$\Delta$$\lambda$ = 5.3$\micron$) over the nights of 2004 March 6, 11, and 12. The chop angle was
121$\degr$ and the instrument position angle was set at 172$\degr$. The nucleus was positioned in the
corner of the array in order to concentrate on possible extended mid-IR emission along the synchrotron
jet emanating from the AGN. On March 6th HD110458 was observed for a PSF and flux calibrator (43 sec
on-source) followed by 6 successive observations of Cen A (each 825 sec on-source) for a total
on-source time of 82.5 minutes on the galaxy. On the following night March 11th, Cen A was observed
twice (each observation 825 sec on-source) followed by a PSF and calibrator star HD 108903 (36 sec
on-source). The same observation sequence was repeated on March 12th.

\section{Results}
In both Epoch A and B we detect low-level extended mid-IR emission near the nucleus of Cen A at
8.8$\micron$, N-band, and 18.3$\micron$. Using multiple observations of the AGN and nearby stellar
targets in Epoch A we determine the central nucleus of Cen A to be unresolved at both 8.8$\micron$ and
18.3$\micron$ and place firm limits on the size based on the FWHM. We make a distinction between the
unresolved nucleus and the surrounding emission and discuss both separately in $\S$ 3.1 and $\S$ 3.2,
followed by analysis of the source of such emission in $\S$ 4.

\subsection{The Central Nucleus}

 Figure 1 shows the FWHM of the
nucleus of Cen A and corresponding PSFs over time. Observations of the galaxy were split up into
nod-sets to match the same integration times of the PSF for careful comparison (65 sec on-source at
8.8$\micron$ and 109 sec on-source at 18.3$\micron$). The overall decrease in FWHM in the 8.8$\micron$
plot corresponds to the lower airmass through the observations as well as an improvement in seeing
throughout the night. A similar plot is shown for 18.3$\micron$ in Figure 1 also. Though there is some
variation in the FWHM, the bright nucleus of Cen A is unresolved. Figure 2 shows azimuthally averaged
radial profiles of the total coadded Cen A data at 8.8$\micron$ and 18.3$\micron$ in comparison to
coadd of all the PSF observations at each wavelength. Given the 3$\sigma$ variation of the PSF it is
clear the bright nucleus of Cen A is unresolved at both wavelengths. Neither the extension claimed by
K03 nor by HKW06 is detected. In the case of K03 it was noted that several hours passed between galaxy
and PSF observations while HKW06 compared individual galaxy exposures of 825 sec on-source with a
single short 43 sec on-source observation of the PSF. As can be seen in Figure 1, the PSF and galaxy
vary significantly over time with changes in seeing and airmass, highlighting the potential dangers in
comparing a single, short PSF measurement to a set of longer galaxy observations.

We agree with the observations of Whysong \& Antonucci (2004) that the nucleus of Cen A is unresolved,
though with multiple PSF observations we are able to place a tighter statistical constraint of the size
of the bright mid-IR nucleus. Following the analysis of Soifer et al. (2000) and Packham et al. (2005),
we suggest that any nuclear extensions would be readily detected at 3 times the standard deviation of
the PSF standard. For our 8.8$\micron$ data the median size of the PSF $\theta$$_{PSF}$=0.3$\arcsec$,
3$\sigma$=0$\farcs$057 ($\theta$$_{sd}$) which excludes the first PSF observation taken under poor
seeing and high airmass that is not representative of the group. For our 18.3$\micron$ data median size
of the PSF $\theta$$_{PSF}$=0.53$\arcsec$, 3$\sigma$=0$\farcs$039 ($\theta$$_{sd}$), based on all 3 PSF
observations. The maximum angular extension $(\theta_{max})$ is defined

\begin{equation}
\theta^{2}_{max} = \theta^{2}_{tot} - \theta^{2}_{PSF}
\end{equation}

\noindent where $\theta$$^{2}_{tot}$ is ($\theta$$_{PSF}$ + $\theta$$_{sd}$)$^{2}$. Thus, any bright
nuclear extended emission must occur on scales $<$0$\farcs$19 ($\lesssim$3.2pc) at 8.8$\micron$ and
$<$0$\farcs$21 ($\lesssim$3.5pc) at 18.3$\micron$ at the level of the FWHM.

\subsection{The Extended Emission}

Although the mid-IR nucleus of Cen A is unresolved when compared to the FWHM of corresponding PSFs
there is clear evidence of surrounding low-level extended emission detected in both Epoch A and B. This
mid-IR emission is faint and only becomes apparent after a 10 pixel (0.89$\arcsec$) gaussian smooth is
applied. Figure 3 clearly shows this emission at 8.8$\micron$ and 18.3$\micron$ from Epoch A overlaid
on Pa-$\alpha$ emission observed by Marconi et al. (2000). Similar structure to that at 8.8$\micron$ is
seen in N-band data from Epoch B (Figure 4). This emission has no effect on the FWHM of the central
nucleus and could not have been the cause of previous detections of a ``resolved" nucleus by K03 or
HKW06. It also is too faint, $>$ 250$\times$ fainter than the peak at both 8.8$\micron$ and N-band of
the 10 pixel smoothed image, to be associated with the 2$\arcsec$ extended emission claimed by
Siebenmorgen et al. (2004) which was only at a level of 1/10th the peak at 10.4$\micron$ (Figure 5).

Low-level extended mid-IR emission is known to exist near the nucleus of Cen A. Observations using the
Infrared Space Observatory (ISO) by Mirabel et al. (1999) and most recently Spitzer observations by
Quillen et al. (2006; 2006b) show the nucleus is surrounded by a dusty warped disk. Observations in
this paper however represent to date the highest resolution mid-IR images that clearly detect this
emission surrounding the nucleus. Analysis of this emission however is limited due to the chop throw of
15$\arcsec$ which results in some chopping onto emission, compromising detailed examination of the flux
and morphological parameters. Though a minor effect with respect to the bright nucleus, the faint
extended emission does show some evidence of negative chop regions. However, the similar mid-IR
structure of this emission at 8.8$\micron$ and N-band despite chopping onto different regions would
seem to show that much of the structure observed is real. In addition the correspondence of the mid-IR
with emission regions of Pa-$\alpha$ detected by Schreier et al. (1998) and Marconi et al. (2000)
reinforces that in fact what we see are primarily real structures (Figures 3 and 4).

\section{Analysis: Sources of emission}
In this section we discuss the possible origins of the mid-IR emission of both the bright unresolved
central nucleus and the low-level extended emission. Due to the the contamination from chopping onto
nearby emission, the analysis of the low-level extended emission is presented in a broad overview.
Additional imaging data resulting in a more detailed analysis of this emission will be put forth by
Leeuw et al. (in preparation).

\subsection{The Nucleus}
\subsubsection{Nucleus: Stellar Activity}
Starburst activity in the nucleus of Cen A has been modeled by Alexander et al. (1999), who fit the
nuclear ($<$ 4$\arcsec$) IR flux $\leq$ 60$\micron$ with approximately equal parts dusty torus and
starburst emission. However, their nuclear starburst component depends strongly on fitting the
11.3$\micron$ PAH feature from Mirabel et al. (1999) that is not seen in the higher spatial resolution
($\sim$ 3$\arcsec$) Siebenmorgen et al. (2004) data. In addition, estimates of the high intrinsic
polarization from near-IR observations by Packham et al. (1996) and Capetti et al. (2000) on order of
11\% to 17\% are consistent with optically thin scattering from a compact torus surrounding the central
engine rather than a starburst.

Combining our data with bolometric luminosity measurements, we can evaluate the likely contribution of
a starburst using a ``luminosity density" analysis similar to that by Soifer et al. (2000, 2001, 2003)
and Evans et al. (2003). They have calculated the surface brightnesses from infrared luminous AGN and
starburst galaxies. This was done by measuring the size or upper limit of the mid-IR size of a compact
region and assuming that this mid-IR region is representative of the infrared emitting region as a
whole. Using the total IR luminosity as an approximation of the bolometric luminosity in these galaxies
and assuming it originates from this region resulted in an estimate of the luminosity density. In
general, they found that galactic HII regions have surface brightnesses ranging from 2 $\times$
10$^{11}$ to 2 $\times$ 10$^{12}$ L$_{\sun}$ kpc$^{-2}$, starburst galaxies range from 2 $\times$
10$^{11}$ to 10$^{13}$ L$_{\sun}$ kpc$^{-2}$, while ULIRGS typically range from 2 $\times$ 10$^{12}$ to
6 $\times$ 10$^{13}$ L$_{\sun}$ kpc$^{-2}$. On small scales ($\leq$ 10 pc) Meurer et al. (1997) found
that star clusters may have a global (UV, IR, and radio) surface brightness as high as 5 $\times$
10$^{13}$ L$_{\sun}$ kpc$^{-2}$. Soifer et al. (2003) found in three Seyferts surface brightnesses on
order of a few times 10$^{14}$ L$_{\sun}$ kpc$^{-2}$, which they argued must be primarily AGN emission,
being well above the luminosity density of bright star clusters and starbursts.

Based on our observations we measure a mid-IR luminosity of the nucleus of Cen A of 1.8 $\times$
10$^{8}$ L$_{\sun}$. This value is similar to the 1.5 $\times$ 10$^{8}$ L$_{\sun}$ value of Whysong \&
Antonucci (2004) based on their 11.7$\micron$ measurement and represents a lower limit to the total
luminosity. Assuming a normal quasar SED Whysong \& Antonucci (2004) estimate the bolometric luminosity
of the AGN to be $\sim$ 2.5 $\times$ 10$^{9}$ L$_{\sun}$. This is similar to the value calculated by
Israel (1998) and within a factor of $\sim$ 2 quoted by Marconi et al. (2001). Given the lower
estimated bolometric luminosity and our maximum size limit of 3.5pc from our 18.3$\micron$ observations
we calculate a luminosity density for Cen A of 2.8 $\times$ 10$^{14}$ L$_{\sun}$ kpc$^{-2}$. These
values are similar to those found by Soifer et al. (2003) for Seyferts in which starburst activity was
considered negligible in regions with a luminosity density $>$ 10$^{14}$ L$_{\sun}$ kpc$^{-2}$. Even a
bright star cluster as suggested by Meurer et al. (1997) (5 $\times$ 10$^{13}$ L$_{\sun}$ kpc$^{-2}$)
could only contribute $<$ 18\% to the overall luminosity density. Thus, this general analysis in
addition to the lack of PAH emission detected by Siebenmorgen et al. (2004) and the strong polarization
observations obtained by Packham et al. (1996) and Capetti et al. (2000) indicate that any starburst
activity in the nucleus of Cen A is negligible compared to the emission from the AGN.

\subsubsection{Nucleus: Narrow Line Region (NLR) Dust} In order to explore the possibility
of central heating of a dusty NLR we calculate color temperatures from the ratio of our 8.8 $\micron$
and 18.3 $\micron$ fluxes (see Radomski et al. 2003). Based on this ratio we calculate a temperature of
the nucleus of $\sim$ 210 K. A similar temperature is obtained if we use the ratio of fluxes at 11.7
$\micron$ and 17.75 $\micron$ from Whysong \& Antonucci (2004). This value is slightly higher than the
$\sim$ 160 K calculated by K03 based on their single 8.8$\micron$ flux. However, we can achieve
approximately the same temperature using their formula for our 8.8$\micron$ measurement so the
difference is entirely due to the calculation method. Taking into account extinction from the dust lane
across Cen A as proposed by K03 of $\tau$ $\sim$ 1 at 10 $\micron$ raises the color temperature by only
20 K to 230 K.

We derive an approximation of the dust grain properties using the methodology applied to NGC 4151
(Radomski et al. 2003) and Circinus (Packham et al. 2005). Assuming a simple uniform dust distribution,
a first-order determination of the size of the region that could be heated by a central source can be
made. Given that dust grains primarily absorb UV-optical radiation and re-emit in the infrared, the
equilibrium temperature of dust in a strong UV field is given by (Sellgren et al. 1983)

\begin{equation}
T\sim \left( \frac{L_{UV}}{16\pi R^{2}\sigma }\frac{Q_{UV}}{Q_{IR}}\right) ^{1/4}
\end{equation}
In the above equation, T is the dust temperature, $L_{UV}$ is the approximate UV-optical luminosity of
the central source, $R$ is the radius from the source in parsecs, $\sigma $ is the Stefan-Boltzman
constant, and $Q_{UV}/Q_{IR}$ is the ratio of the Planck averaged UV absorption coefficient to the
infrared emission coefficient. Values of $Q_{UV}/Q_{IR}$ are dependent on the dust grain size and
composition and are given by Draine \& Lee (1984), Laor \& Draine (1993), and Weingartner \& Draine
(2001) for graphite and ``smoothed astronomical'' (SA) silicate.

Using the bolometric luminosity of 2.5 $\times$ 10$^{9}$ L$_{\sun}$ as an estimate and a dust
temperature of 200 K, Cen A could heat NLR dust similar to that in Circinus and NGC 4151
(0.003-0.01$\micron$ silicate; 0.015-0.04$\micron$ graphite) up to a radial distance of 9 - 14pc
($\sim$0$\farcs$5 - 0$\farcs$8) from the the central engine. This is much larger than the maximum size
of the unresolved mid-IR nucleus (0$\farcs$21; 3.5 pc diameter) and shows that the central AGN is
capable of heating all the dust in the core. With these observations it is difficult to determine how
much of this unresolved emission may be distributed between a compact NLR and torus. However, if we
assume a similar scenario as in NGC 4151 then the NLR may only contribute $\sim$ 10-20\% of the mid-IR
emission (Radomski et al. 2003; Groves, Dopita, \& Sutherland 2006).

\subsubsection{Nucleus: Dusty Torus} The maximum size we obtain for the mid-IR
emission from dust associated with a compact geometrically and optically thick torus at the level of
the FWHM is $<$0$\farcs$19 ($\lesssim$3.2pc) at 8.8$\micron$ and $<$0$\farcs$21 ($\lesssim$3.5pc) at
18.3$\micron$. This is consistent with observations at K-band (2.2$\micron$) FWHM $<$ 0.2$\arcsec$
($<$3.4pc) (Schreier et al. 1998) as well as the polarization observations of Capetti et al. (2000),
which suggest scattering off a 2pc disk. It is also close to the 4.07pc x 3.05pc (scaled to 3.5 Mpc)
torus modeled by Alexander et al. (1999), although as discussed above their starburst fit may not be
accurate if there is no nuclear PAH emission.

Applying the recent models of Nenkova et al. (2002), we in fact predict that the mid-IR emission of the
Cen A torus is unresolvable in the present observations. These models are far more robust than our
earlier calculations of dust in the relatively optically thin NLR, taking into account radiative
transfer and cloud shadowing necessary in the optically thick environment of a torus. In these models
the inner radius of the torus scales with the square root of the AGN luminosity and inversely with
temperature to the 5/2 power.  At the dust sublimation temperature (1200 K), the inner radius of the
Cen A torus may be only 0.07 pc. While the outer edge of the torus may be large ($\sim 5$ pc), in these
inhomogeneous torus models the cloud distribution declines rapidly ($\propto r^{-q}$, where $q = 1$, 2,
or 3).  Thus, most of the clouds are strongly concentrated within the inner parsec (Figure 6). Also
though the hottest dust has $T =1200 $K, the contributions of the many cooler indirectly-heated clouds
are significant, and the net spectrum of the clumpy torus appears much cooler.  For the models
presented above, $F_\nu$ peaks around 25 $\mu$m, with the color temperature of the
8.8$\micron$/18.3$\micron$ flux density ratio indicating $T\approx 200$K.

The model mid-IR emission does not always trace the cloud distribution. Because of the large optical
depths through the densest regions of the torus, the heating and subsequent emission can be suppressed
in the torus midplane. These effects are strongest at shorter wavelengths and for less compact cloud
distributions, where the strongest observable emission extends along the torus axis.  In such
instances, some hot, luminous material (on the projected front or rear of the torus) is not completely
obscured. We modeled a range of torus parameters, which include the radial distribution, $q$, the
optical depth per cloud measured in the $V$ band, $\tau_V$, and the total number of clouds through the
equatorial plane, $N_C$. We show the results for a model that fits the observed 10$\micron$ nuclear
spectrum of NGC 1068 well (Mason et al. 2006), with $q = 2$ and $N_C = 8$, and a comparison model in
which $q = 1$ and $N_C = 5$. In both models, $\tau_V = 40$, the outer radius is 30 times the dust
sublimation radius, and the cloud distribution is gaussian in elevation above the torus mid-plane,
without a sharp cutoff. These two sets of parameters yield cloud distributions and images that are
characteristic of the full parameter space we explored.

The torus is viewed edge-on in the simulated images at 8.8 and 18.3$\micron$ (Figure 6).  The
18.3$\micron$ emission generally traces the cloud distribution because optical depth effects are not
significant, both for heating the clouds and their subsequent emission.  Directly-heated clouds that
are not blocked along the line of sight dominate the emission at 8.8$\micron$.  The resulting images
are extended perpendicular to the cloud distribution, and in the less compact ($q=1$) distribution,
this extension persists on larger scales.  The exact emission profile does depend on the model
parameters, but in all cases the strong extended emission (which the FWHM of the simulated images
themselves indicates) is confined to scales of $0\farcs02$. In the simulations, the flux declines to
10\% of the peak strength on scales of $0\farcs08$ or less. In general, the emission is more compact in
the steeper cloud distributions, and it is always much smaller than the $0\farcs2$ upper limit we
measure in the data. This approaches the scales of HST STIS observations of Marconi et al. (2006) which
indicate the central dark object in Cen A is no more than 0$\farcs$036-0$\farcs$04 (0.6-0.7pc) in
radius. Thus, careful interferometry will be required to resolve this structure.

\subsubsection{Nucleus: Synchrotron Emission}
As in the case of M87 (Perlman et al. 2001), the powerful radio jet of Cen A could provide a
significant fraction of the mid-IR emission through synchrotron radiation. Chiaberge et al. (2001)
claim the entire nuclear SED of Cen A can be fit with a synchrotron self-Compton (SSC) model. Also
recent mid-IR interferometry models by Meisenheimer et al. (2007) suggest a tiny predominantly SSC core
($<$0.2pc). They model that the SSC core contributes between 80-60\% of the emission between 8 and 13
$\micron$ respectively. However there are several problems with this scenario.

First, the lack of polarization at  $\sim$ 1000$\micron$ (Packham et al. 1996) is difficult to
reconcile with a predominantly synchrotron core. Second, Meisenheimer et al. (2007) reference the
variability seen at L-band ($\sim$3.5$\micron$) by Lepine et al. (1984) detailing an increase in flux
by a factor of 5 between 1971-1981 as evidence of the variable synchrotron nature of the IR core. If
the mid-IR core is dominated by synchrotron emission then similar variability should be seen at $\sim$
10$\micron$. However mid-IR (10$\micron$) measurements taken coincident with the L-band observations
(Becklin et al. 1971, Kleinmann et al. 1974, Grasdalen et al. 1976) seem inconsistent with variability
trends seen at X-ray and radio wavelengths (Beall et al. 1978; Abraham et al. 1982). Indeed Telesco
(1978) comparing mid-IR (10$\micron$) observations from 1971-1978 concluded that there was no evidence
of mid-IR variability during this period. Comparing mid-IR observations taken from 1984 to 2004 also
show no significant variation at $\sim$10$\micron$ after accounting for filter and calibration
differences (Alexander et al. 1999; Mirabel et al. 1999; Krabbe et al. 2001; Siebenmorgen et al. 2004;
Karovska et al. 2003; Weedman et al. 2005; Hardcastle et al. 2006; this paper).  And finally, the
fluxes measured by Meisenheimer et al. (2007) for the SSC core at 11.4$\micron$ are a factor of $\sim$
4 smaller that the unresolved core measured by Whysong \& Antonucci (2004) at 11.7$\micron$. Given that
the high spatial resolution Siebenmorgen et al. (2004) spectra show no PAH emission it is unlikely that
this could affect the flux. Thus at least at the longer portion of the 10$\micron$ window it appears
that there is ample excess emission possibly from a torus that could dominate the mid-IR on scales $<$
0$\farcs$3.

Another method to estimate the synchrotron emission in the nucleus of Cen A is to use a simple power
law model such as that used for M87 (Perlman et al. 2001). A strong linear correlation between the
optical and radio core values of FR I galaxies found by Chiaberge, Capetti, and Celotti (1999) hinted
at a common synchrotron origin for this emission. Capetti et al. (2000) used this correlation and the
radio core flux of Cen A of 9.1 Jy at 15GHz (Clarke, Burns, \& Norman 1992) to predict the synchrotron
emission at 2$\micron$ of 10mJy. This trend also follows closely with the synchrotron models of HKW06
which fit the non-nuclear ``inner", ``middle" and ``outer" lobes of the radio jet in Cen A. Assuming
this trend and a radio core flux of 9.1Jy we predict a nonthermal mid-IR flux density between 8 and
18$\micron$ of 30 to 50mJy; more than an order of magnitude less than what is measured. This may
indicate that synchrotron emission in the nucleus of Cen A is negligible in the mid-IR in comparison to
emission from dust.

%********************************************************************

\subsection{The Extended Emission}
\subsubsection{Extended: Stellar Activity}
The nucleus of Cen A is most likely dominated by the AGN, however much of the surrounding extended
emission is likely due to stars. The close correspondence between much of this emission
$\gtrsim$5$\arcsec$ from the nucleus at 8.8$\micron$ and N-band and the Pa-$\alpha$ data from Marconi
et al. (2000) (Figures 3 and 4) indicate much of what we are seeing is associated with regions of star
formation. The same has been seen in other extragalactic star forming regions by Alonso-Herrero et al.
(2006). In general, these clumps as well as much of the emission outline the twin parallel lanes of
emission seen as part of the large scale warped disk in the lower resolution Spitzer mid-IR data
(Quillen et al. 2006; Figure 1). At 8.8$\micron$ and N-band the Pa-$\alpha$ clumps C, D, E, F, and G
seen by Schreier et al. (1998) are all detected along the direction of the top lane of emission seen by
Spitzer. Along the bottom lane only the Pa-$\alpha$ clump K is detected.  Several other mid-IR sources
of emission associated with regions of Pa-$\alpha$ emission are also obvious in Figures 3 and 4. At
18.3$\micron$ little evidence of this emission is seen. This is most likely due to two reasons: (1) the
lower sensitivity of the 18$\micron$ (Qa) filter and (2) star formation regions typically have a PAH
component at 8.6$\micron$ and 11.3$\micron$ which is readily detected in the 8.8$\micron$ and N-band
filters. However, at both 8.8$\micron$ and N-band some Pa-$\alpha$ clumps are devoid of mid-IR
emission. Unfortunately this may be due to the limited chop throw (15$\arcsec$) of the data discussed
previously which is certainly chopping onto other regions of low level emission creating negative areas
in the field of view. Thus extensive analysis of theses star formation regions is not feasible at this
time. However, there is an overall good correspondence between the Pa-$\alpha$ and mid-IR.

\subsubsection{Extended: Narrow Line Region (NLR) Dust}
Although the nucleus can be heated entirely by the AGN it is unlikely that the bulk of the extended
emission could be heated by the central engine. This low level-extended mid-IR emission, in some cases
up to 10$\arcsec$ (170pc) away from the nucleus would have to be extremely cool on order of $\sim$ 60K
assuming the same simple formula as in $\S$ 4.1.2. Assuming a normal blackbody SED of dust this cool
however would cause significant emission (on order of a few Jy) at longer wavelengths such as
18.3$\micron$ where no significant extended emission is seen on scales $>$3$\arcsec$. This provides
further evidence that much of this mid-IR emission, especially that coincident with the Pa-$\alpha$
emission, is associated with local stellar activity.

The only significant 18.3$\micron$ emission detected is a small clump of emission at a distance of
2.8$\arcsec$ (48pc) from the nucleus (PA=64$\degr$; measured North through East) approximately
coincident with Pa-$\alpha$ clump A from Schreier et al. (1998) (see Figure 3). Krajnovi{\'c}, Sharp,
\& Thatte (2006) suggest this clump as well as clump B may be due to jet induced star formation.
Emission is detected in both clumps at 8.8$\micron$ while there is a $\sim$2$\sigma$ detection of clump
B at 18.3$\micron$. If this mid-IR emission is truly associated with shock induced star formation
regions then we would expect heating of the surrounding dust to be local.

We measure the the flux density in a 2$\arcsec$ diameter area near clump A to be $\sim$ 4mJy at
8.8$\micron$ and $\sim$ 40mJy at 18$\micron$, given the possibility of emission being chopped onto
these can be considered rough estimates. Near clump B the mid-IR emission is similar to that of clump A
$\sim$ 4mJy at 8.8$\micron$ with very little emission ($\sim$ 17mJy) at 18.3$\micron$. This results in
an optically thin estimate of the temperature of 150-200K near clump A and even warmer near clump B
which is farther away at 3.5$\arcsec$ (60pc). Given that we estimate that 14pc is approximately the
maximum distance that the AGN (2.5 $\times$ 10$^{9}$ L$_{\sun}$) can heat dust to 200K this would rule
out mid-IR emission near clump B as being centrally heated dust. Even for clump A at a temperature of
150K only the smallest grains of graphite (0.003-0.005um) could be heated by the central engine. Though
these are very simple estimates it does imply that local heating of dust due to shock induced star
formation associated with the radio ejecta is a possible explanation for mid-IR emission near clumps A
and B. However, if there is an enhancement in luminosity in the direction of the NLR along the radio
axis (NE of the nucleus), as would be the case if the ionizing luminosity is anisotropic, then it may
be possible to heat dust out to a radius consistent with clumps A and B. This could also explain the
faint mid-IR extension seen on the opposite side (SW of the nucleus) at 8.8$\micron$. This is thought
to be the case in NGC 4151 (Radomski et al. 2003) where extended mid-IR emission on either side of the
nucleus along the radio axis is thought to arise from a dusty NLR within the ionization cone heated by
the central engine. In this case the NLR of NGC 4151 may see an ionizing luminosity a factor of
$\sim$13$\times$ greater than that seen from Earth (Penston et al. 1990). Anisotropies and beaming
along the NLR has been predicted as strong as 200$\times$ in some galaxies (Baldwin, Wilson \& Whittle
1987). In Cen A a beaming factor or underestimate of the true luminosity of 10-20$\times$ would be
needed to be consistent with the central heating of clumps A and B within an ionization cone.

\subsubsection{Extended: Dusty Torus} CO and H$_{2}$ measurements of Cen A show a nuclear ring of outer radius 80-140pc,
with an inner hole of diameter 40pc enclosing a mass of 10$^{9}$ L$_{\sun}$ (Israel et al 1990, 1991;
Rydbeck et al. 1993; Marconi et al. 2001). This ring is aligned perpendicular to the radio jet and has
been speculated to be associated with the very outer parts of a torus. This would seem to contradict
our earlier results which indicate a highly compact torus $<$3.5pc. The difference however is entirely
due to what is defined as the ``torus" in AGN. Shi et al. (2006) describes the ``torus" as a
multilayered system with and ``inner" ($<$0.1pc) ``middle" (0.1-10pc) and ``outer" (10-300pc)
structure. In this scenario the compact torii seen in the mid-IR would typically correspond to the
``middle" disk while the CO/H$_{2}$ is associated with the ``outer" disk. Similar to Cen A, NGC 1068
also contains a compact mid-IR torus $<$15pc (Mason et al. 2006) surrounded by a large 100pc scale CO
disk. Mid-IR polarization observations of NGC 1068 by Packham et al. (2007) provide continuity between
these structures suggesting that the compact geometrically and optically thick torus is often
surrounded by a larger and more diffuse structure associated with the dusty central regions of the host
galaxy. Other dusty structures such as nuclear galactic bars may further merge into this scenario.
Observations by Marconi et al. (2001b) even suggest that the H$_{2}$ structure shows kinematical
evidence consistent with a galactic bar fueling the AGN, though they cannot entirely rule out a
disk/torus structure.

Possible evidence for mid-IR emission associated with this ``outer" torus/nuclear bar can be seen at
both 8.8$\micron$ and N-band (Figures 3 and 4). At both wavelengths there appears to be a ridge of
emission extending up from the NW of the nucleus towards Pa-$\alpha$ clump G. At 8.8$\micron$ some
mid-IR emission is also seen on the SE side. In all cases this emission seems devoid of corresponding
Pa-$\alpha$ emission associated with stellar activity as discussed earlier. Given that the data at
8.8um and N-band is chopping on different areas of the galaxy (see Observations $\S$ 2) it is unlikely
that at least the NW structure seen at both wavelengths is excessively distorted due to improper chop
subtraction. This structure possibly indicates that the emission detected in CO/H$_{2}$ contains a
dusty component. However, using simple calculations from $\S$ 4.1.2, even dust as small as
0.003$\micron$ at this distance (5-7$\arcsec$) would have to be on order of 70-80K to be heated by the
central AGN. This is not reasonable as such cool dust should emit substantially more (on order of a few
Jy) at 18$\micron$ where no significant emission is detected along this ridge. A more likely
possibility is this emission is due to PAH emission as detected by ISO and Spitzer at least at
8.8$\micron$ and possibility also 11.3$\micron$ in the N-band.

NICMOS imaging by Schreier et al. (1998) shows a 1$\arcsec$x2$\arcsec$ (20pc radius) emission line
region interpreted as a gaseous disk also possibly associated with a multi-layered torus which is
nearly parallel to the radio jet and perpendicular to the CO disk. Schreier et al. (1998) hypothesized
this peculiar angle may be caused by one of two scenarios. Either that it had formed recently enough
(possibly associated with a merger or infall event) such that it had not time to become aligned with
the black hole spin axis, or because it was along the major axis of the bulge that it was dominated by
the galaxy gravitational potential rather than that of the black hole. In either case these
observations indicate that there may be an outer gaseous ring of material that is highly warped with
respect to the inner compact mid-IR torus. Though we do generally detect low-level mid-IR emission
along this direction we detect no extension associated with this structure at the level of the FWHM.

\subsubsection{Extended: Synchrotron Emission}
As discussed earlier, in both Epoch A and B data extended mid-IR emission is detected approximately
coincident with Pa-$\alpha$ clumps A and B from Schreier et al. (1998). This emission also falls very
near the axis of the powerful radio jet in Cen A. At N-band especially (Figure 4) emission is seen
along the inner part of the radio jet extending out past the Pa-$\alpha$ clumps A and B. As discussed
above it is difficult to rule out shock induced star formation and/or dusty NLR emission. Furthermore
due to the possibility of inadequate chop-subtraction attempting to fit a synchrotron model to evaluate
its contribution, if any, is difficult. Thus we cannot confirm nor rule out some contribution from
synchrotron radiation emanating from the jet near the nucleus. Farther out along the jet axis in the NE
direction negative emission from improper chop subtraction becomes more apparent in both Epoch A and B
data and some faint mid-IR emission associated with the jet is most likely lost. However in Epoch B at
N-band there is clear evidence of mid-IR emission located farther out and coincident with the radio
axis. This new source is denoted as RPL 1 (Figure 4).

RPL 1 is located in a very bright area of the radio jet at a distance of 18.4$\arcsec$ PA=(53$\degr$
measured North to East). It is outside the field of view of the Epoch A data and thus cannot be
characterized at 8.8$\micron$ and 18.3$\micron$. Given the distance as projected on sky 18.4$\arcsec$
is equal to $>$300 pc. This rules out dust heated by the central source, as it would take a beaming
factor of $>$500 to heat even the smallest dust grains to a temperature of 200K. If no beaming is
assumed the maximum temperature that a blackbody heated by the central AGN could obtain at that
distance is 42K, which given an approximate flux density of 2mJy at N-band would require a 25$\micron$
flux several hundred thousand times what is measured by IRAS. Confidence in the detection of this
source is strong as it is seen on all three nights observed at N-band and after a 20 pixel
(1.78$\arcsec$) offset/dither between nights. Stacking all three nights of data results in a
3-4$\sigma$ detection (Figure 4). Mid-IR emission from a synchrotron jet has been seen in M87 (Perlman
et al. 2001) and at larger radii from Cen A by Spitzer (HKW06), however in these cases the mid-IR
morphology and position showed a strong spatial and morphological correspondence to that of the radio
emission from the jet. This does not seem to be the case in RPL 1 which is located offset
$\sim$2$\arcsec$ from a bright radio knot and with a much less elongated shape. Another possible
explanation is a dusty clump heated by the impact or shock of the outflow impinging on surrounding
material associated with jet induced star formation. However it is possible that the morphology of this
weak source is affected by improper chop subtraction of the crowded mid-IR emission in Cen A and thus
it requires further investigation.

\section{Conclusions}

We detect a bright unresolved mid-IR nucleus surrounded by low-level emission in the central region of
Centaurus A. We find the bright central nucleus of Cen A to be unresolved in the mid-IR at the level of
the FWHM. Using multiple PSF measurements and their variation, we place an upper limit on the size of
the mid-IR nucleus of Cen A of 3.2 pc (0$\farcs$19) at 8.8$\micron$ and 3.5 pc (0$\farcs$21) at
18.3$\micron$. This is consistent with our models which predict the mid-IR emission from a torus in Cen
A to be unresolved with a size of 0$\farcs$05 or less. The primary source of the nuclear mid-IR
emission is likely to be associated with this dusty torus and NLR emission with minimal contributions
from starburst and synchrotron emission.

Extended mid-IR emission in Cen A is generally coincident with Pa-$\alpha$ regions most likely due to
stellar activity. A ridge of mid-IR emission perpendicular to the radio axis and devoid of much
Pa-$\alpha$ emission may be associated with a CO, H$_{2}$ outer ring/bar. Emission associated with
Pa-$\alpha$ clumps A and B from Schreier et al. (1998) could be due to shock heating or possibly
centrally heated dust in a NLR if anisotropic beaming of radiation along the radio axis is taken into
account. A new mid-IR source RPL 1, is detected along the radio axis at a distance of 18.4$\arcsec$
(PA=53$\degr$) and is possibly due to shock heating of material impacted by the jet or synchrotron
emission. Overall the mid-IR emission in the nuclear region is shown to be very complex and in need of
further study.

\section{Acknowledgments}
JTR, JMD, and RM acknowledge the support of the NSF based on observations obtained at the Gemini
Observatory, which is operated by the Association of Universities for Research in Astronomy, Inc.,
under a cooperative agreement with the NSF on behalf of the Gemini partnership: the National Science
Foundation (United States), the Particle Physics and Astronomy Research Council (United Kingdom), the
National Research Council (Canada), CONICYT (Chile), the Australian Research Council (Australia), CNPq
(Brazil), and CONICET (Argentina). We also thank Matthew Sirocky for his assistance with the torus
simulations and NAL acknowledges support from NSF award AST-0237291. JTR would also like to acknowledge
Craig Markwardt for his very useful IDL routines which were used to fit Moffat profiles and Alessandro
Marconi for trying to find his original Pa-$\alpha$ images. CP would like to acknowledge NSF grant
0206617. All authors would also like to thank the referee for the helpful comments that improved this
paper.

\clearpage
\begin{landscape}
\begin{deluxetable}{lccccc}
\tabletypesize{\scriptsize} \tablecaption{Cen A Flux Density Measurements. \label{tbl-1}}
\tablewidth{0pt} \tablehead{ \colhead{Reference} &\colhead{Wavelength ($\micron$)} &\colhead{Telescope}
&\colhead{Resolution} & \colhead{Flux Density (Jy)} &\colhead{Time On-Source (sec)}} \startdata
This Paper  & 8.8 & Gemini South (8m)& 0$\farcs$30&  0.71$\pm 0.04$ \tablenotemark{a} & 2000 \\
This Paper  & 18.3 & Gemini South (8m)& 0$\farcs$53 &  2.63$\pm 0.65$ \tablenotemark{a} & 1550  \\
\cutinhead{Previous Observations}
Karovska et al. (2003) & 8.8 & Magellan (6m) & 0$\farcs$45 &  0.9$\pm 0.1$  & 1400 \\
Karovska et al. (2003) & 10 (N-band)  & Magellan (6m) & 0$\farcs$45 &  1.4$\pm 0.5$  & 900\\
Siebenmorgen et al. (2004) & 10.4 & ESO (3.6m) & 0$\farcs$5 &  0.65$\pm 0.065$  & 645\\
Krabbe et al. (2001) & 10 (N-band) & ESO MPI (2.2m) & 1$\arcsec$ &  0.75$\pm 0.055$ & 600 \\
Hardcastle, Kraft, and Worrall (2006) & 10 (N-band) & Gemini South (8m) & 0.4$\arcsec$ &  1.1$\pm 0.11$ & 8250\\
Whysong \& Antonucci (2004) &11.7 & Keck I (10m)& $\sim$0$\farcs$3 &  1.6$\pm 0.2$  & -\\
Whysong \& Antonucci (2004) &SiC ($\sim$11.7) & Keck I
(10m) & $\sim$0$\farcs$3 & 1.8$\pm 0.2$  & - \\
Whysong \& Antonucci (2004) &17.75 & Keck I (10m) & $\sim$0$\farcs$5 &  2.3$\pm 0.2$  & -\\

\enddata

\tablenotetext{a}{Errors in flux density are dominated by uncertainty in calibration
 ($\leq$ 5\% at 8.8$\micron$ and $\leq$ 25\% at 18.3$\micron$) but also include a small statistical error based on the aperture size.}

\end{deluxetable}
\end{landscape}

\begin{figure}
\epsscale{0.6} \plotone{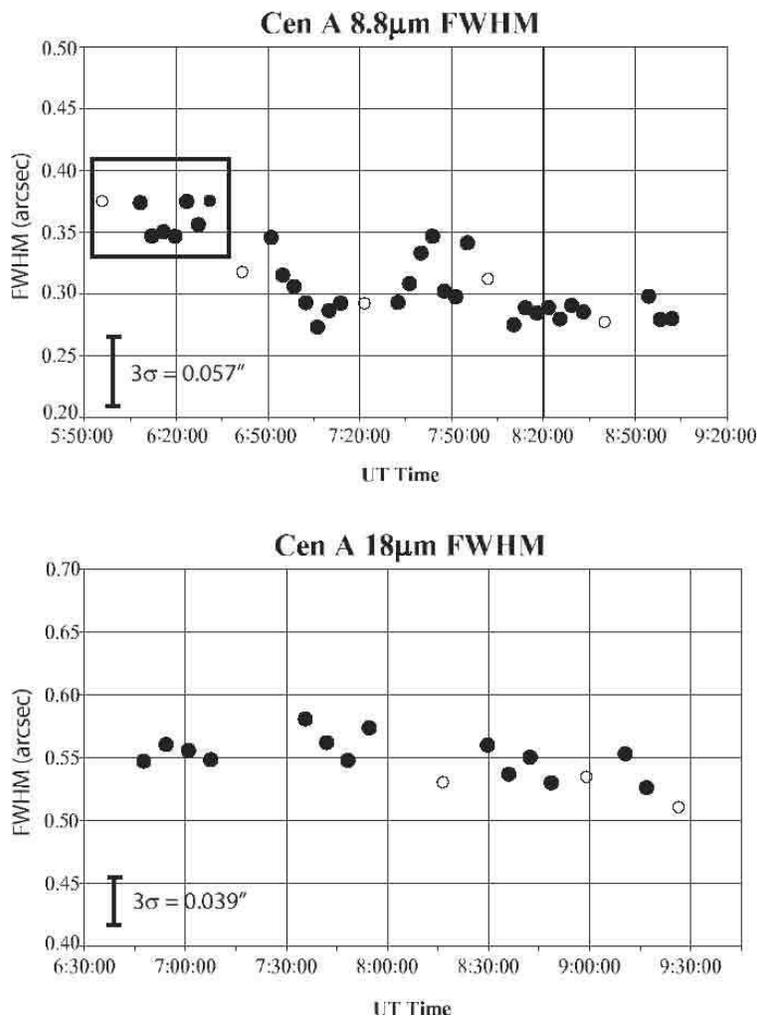} \caption{This figure shows the FWHM of Cen A (filled circles) in
comparison to that of the PSF (open circles) vs. the time observed each night for 8.8$\micron$ and
18.3$\micron$. Each circle corresponds to equal time intervals for galaxy and PSF (65 sec on-source at
8.8$\micron$ and 109 sec on-source at 18.3$\micron$). The box in the 8.8$\micron$ figure indicates data
taken at high airmass and relatively poor seeing that was not used in the overall analysis. In some
cases the galaxy observations contained slightly more time left over after being divided into intervals
equal to that of the PSF ($<$22 sec on-source). In these cases the data was not plotted as it was of
short enough duration and did not add any new information to the analysis.}
\end{figure}

\begin{figure}
\epsscale{1.0} \plotone{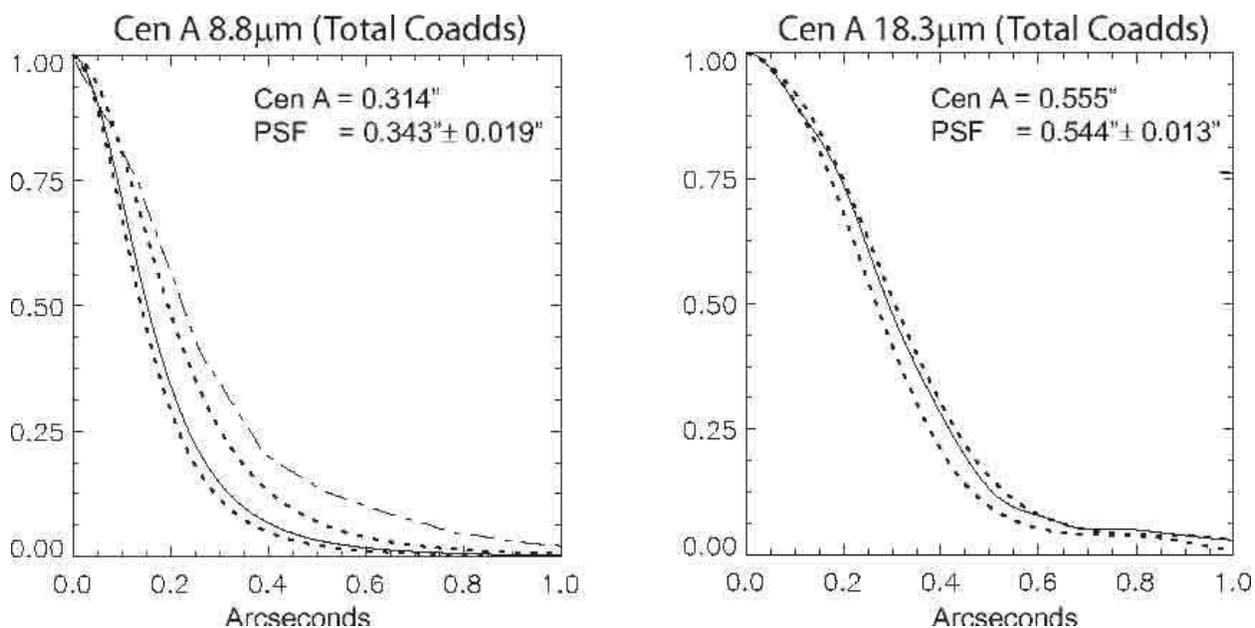} \caption{Azimuthally averaged radial profiles of Cen A in comparison
to the PSF. Solid lines represent the galaxy while dotted lines represent the 3$\sigma$ upper and lower
limit in the PSF FWHM size based on the total standard deviation of the PSFs at each wavelength. These
images represent a comparison of the fully stacked galaxy data (33.3 minutes on-source at 8.8$\micron$
; 25.8 min on-source for 18.3$\micron$) compared to a stack of all PSFs that were taken between galaxy
observations. The dash dot lines represents the measurement of Cen A at 8.8$\micron$ from Figure 2 of
Karovska et al. (2003). Our 3$\sigma$ upper on the PSF is very close to their single PSF measurement.
Thus their 0$\farcs$17 size measurement is likely due to seeing.}
\end{figure}

\begin{landscape}
\begin{figure}
\epsscale{1.0} \plotone{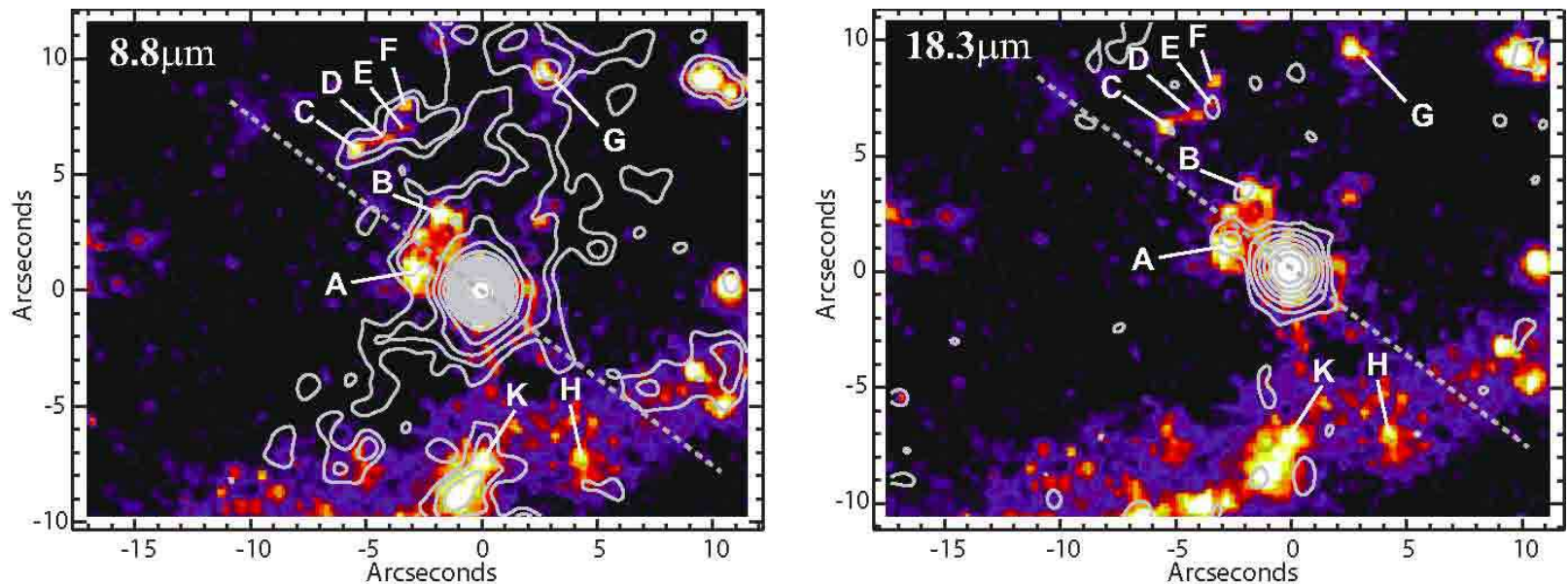} \caption{Grey contours represents our heavily smoothed (10 pixel
gaussian=0.89$\arcsec$) mid-IR data at 8.8$\micron$ (left) and 18.3$\micron$ (right). Contours are
logarithmic with a factor of 1.66 (8.8$\micron$) and 2.11 (18.3$\micron$) between them. The lowest
contours represent 2$\sigma$ above the smoothed background (0.006mJy/pixel at 8.8$\micron$ and
0.066mJy/pixel at 18.3$\micron$). Color images are Pa-$\alpha$ emission from HST (Marconi et al. 2000).
Letters represent the bright Pa-$\alpha$ emission regions detected by Schreier et al. (1998). The
dotted line represents the approximate angle of the radio jet axis (PA=53$\degr$). }
\end{figure}
\end{landscape}

\begin{landscape}
\begin{figure}
\epsscale{1.0} \plotone{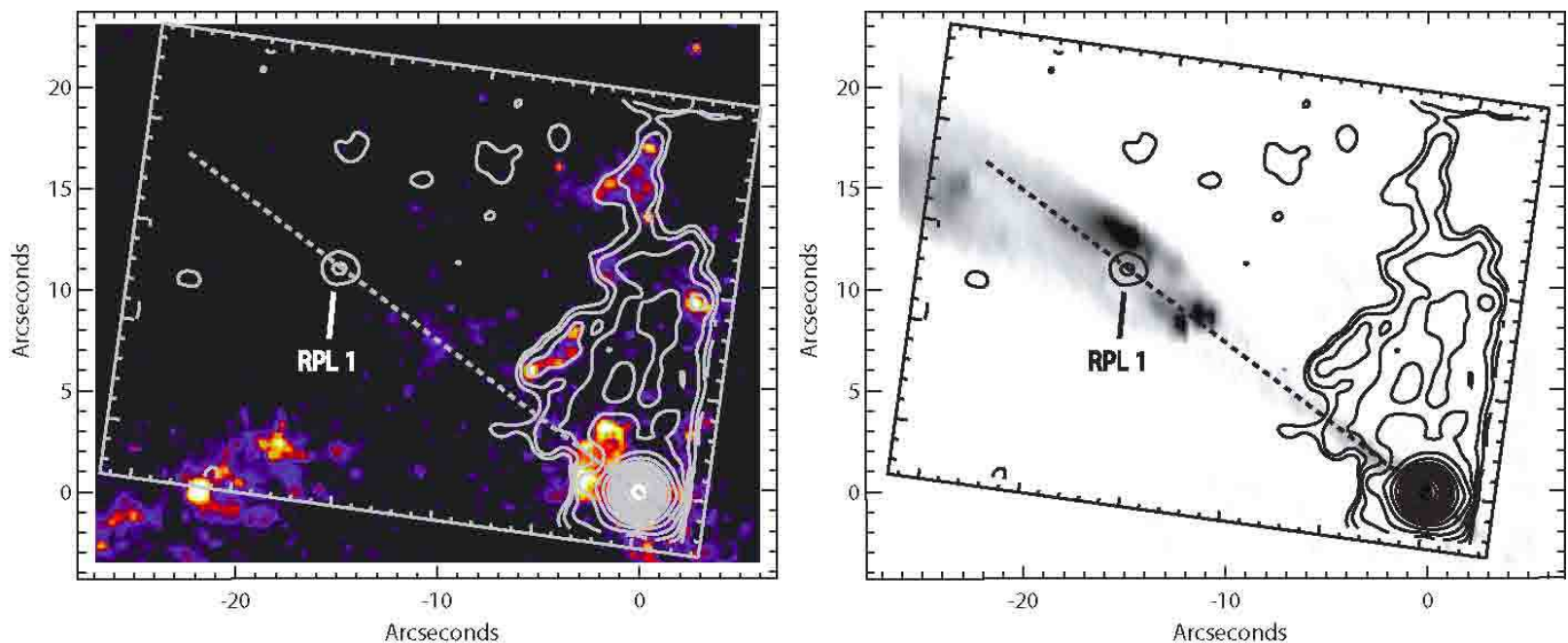} \caption{Contours represent a stack of N-band data from all three
nights (March 6, 11, and 12, 2004) retrieved from the GSA archive. Contours are logarithmic with a
factor of 1.79 between them. The lowest contours represent 2$\sigma$ above the smoothed background
(0.0034 mJy/pixel). The left image shows the contours overlayed on the Pa-$\alpha$ map from Marconi et
al. (2000) as in Figure 3. The right image shows the contours overlayed on the greyscale radio map of
Hardcastle et al. (2003). The mid-IR source RLP1 is detected coincident with the approximate radio jet
PA=53$\degr$ (dotted line).}
\end{figure}
\end{landscape}

\begin{figure}
\epsscale{0.6} \plotone{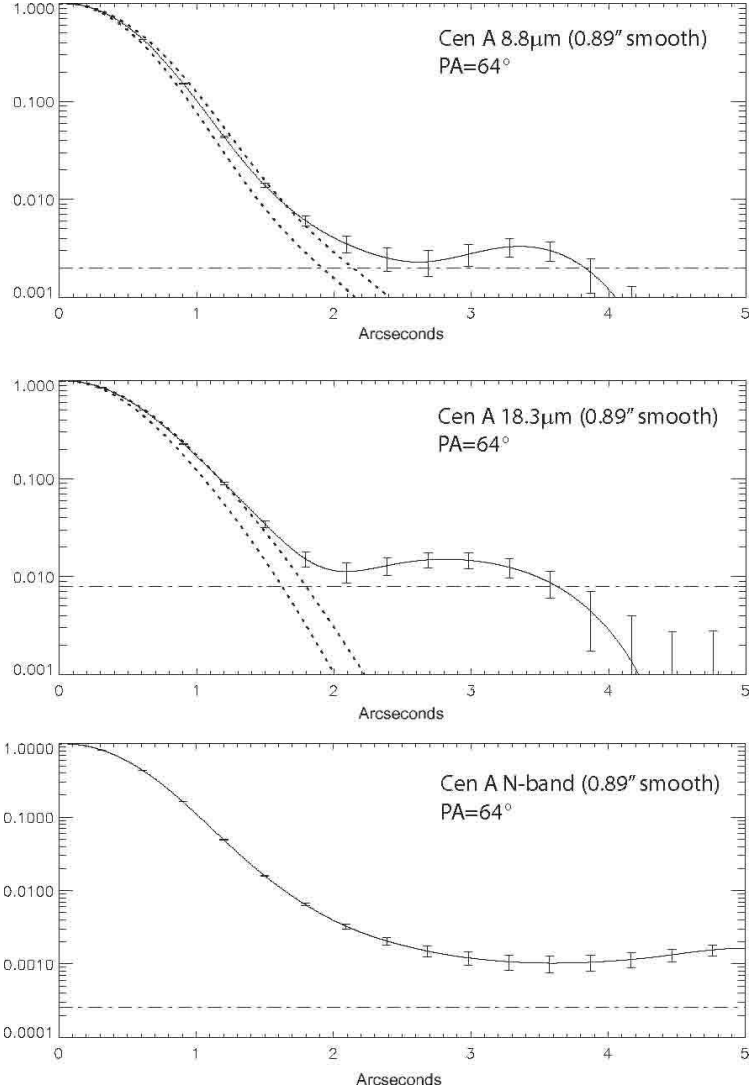} \caption{Line cuts of Cen A at a PA=64$\degr$ after a 10 pixel
(0.89$\arcsec$) Gaussian smooth. Solid and dotted lines represent the same as Figure 2. The dash-dot
line represents 3$\sigma$ above the background and error bars represent $\pm$$\sigma$ of the
background. The position angle is along the extended emission clump detected at 18.3$\micron$ near
Pa-$\alpha$ source A from Schreier et al. (1998) and close to the jet axis. Low level emission near
this clump is clearly detected at 8.8$\micron$, 18.3$\micron$, and N-band greater than 3$\sigma$ above
the background.}
\end{figure}

\begin{figure}
\epsscale{1.0} \plotone{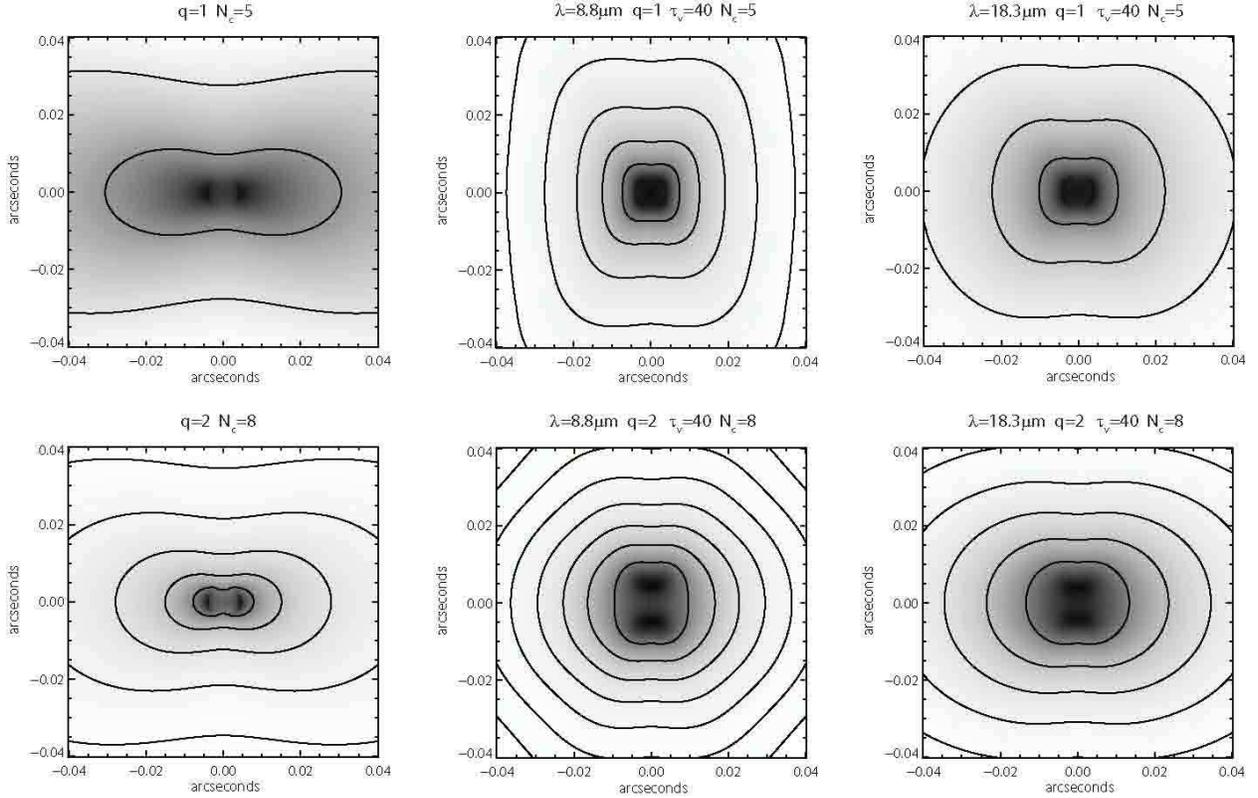} \caption{Number of clouds integrated along the line of sight, for
radial distribution $q = 1$ (top left), and $q = 2$ (bottom left). For the total number of clouds along
an equatorial ray to the central engine, $N_C = 5$ (top left) and 8 (bottom left), the maximum number
of clouds along the line of sight is 12 and 25, respectively, for maximum integrated optical depth of
480 and 1000 in the V band. Beside these cloud distributions are simulated images of emission at 8.8
(center panels) and 18.3 $\mu$m (rightmost panels). In these models, the optical depth per cloud
$\tau_V = 40$ and the outer radius of the cloud distribution is located at 30 times the dust
sublimation radius.  The 18.3 $\mu$m images of both models are similar to each other and to the cloud
distribution because both the heating and subsequent emission are insensitive to optical depth effects.
In contrast, the 8.8$\mu$m emission does not trace the cloud distribution because of the large optical
depth of the torus and the strong temperature dependence, especially when the distribution is not
compact ($q=1$).  Here the emission extends along the polar axis, which is the location of clouds that
directly view the AGN and that are not blocked along the line of sight.  In all cases, the emission is
confined to scales much smaller than spatial resolution of our observations. The images are scaled
linearly, and the contours are logarithmically spaced, beginning at 50\% of the peak value and
declining by factors of two.}
\end{figure}

\end{document}